\documentclass[%
 aip,
rsi,%
amsmath,amssymb,
preprint,%
]{revtex4-1}

\usepackage{times}
\usepackage{tgtermes}
\usepackage[T1]{fontenc}
\usepackage[applemac]{inputenc}
\usepackage{setspace}
\usepackage{graphicx}
\usepackage{subfigure}
\usepackage{xcolor}

\newcommand{\solid}{\emph{solid}}
\newcommand{\liquid}{\emph{liquid}}

\begin{document}

\begin{center}
\begin{spacing}{2.05}
{\fontsize{20}{20}
\bf
Sedimentation of a suspension of paramagnetic particles in an external magnetic field}
\end{spacing}
\end{center}
\vspace{-1.25cm}
\begin{center}
{\fontsize{14}{20}
\bf
J. VESSAIRE, N. PLIHON, R. VOLK, M. BOURGOIN\\
\bigskip
}
{\fontsize{12}{20}
Univ Lyon, ENS de Lyon, Univ Claude Bernard, CNRS, Laboratoire de Physique, F-69342 Lyon, France\\
}
\end{center}

\vspace{10pt}

{\fontsize{16}{20}
\bf
Abstract : 
}
\bigskip

\textit{
We investigate the sedimentation of initially packed paramagnetic particles in presence of a homogeneous external magnetic field, in a Hele-Shaw cell filled with water. Although the magnetic susceptibility of the particles is small and the particle-particle induced magnetic interactions are significantly smaller compared to the gravitational acceleration, we do observe a measurable reduction of the decompaction rate as the amplitude of the applied magnetic field is increased. While induced magnetic dipole-dipole interactions between particles can be either attracting or repulsive depending on the particles relative alignment, our observations reveal an effective overall enhancement of the cohesion of the initial pack of particles due to the induced interactions, very likely promoting internal chain forces in the initial pack of particles. The influence of the magnetic field on the particles once they disperse after being decompacted is on the other hand found to remain marginal.}

\vspace{28pt}

{\fontsize{14}{20}
}

\bigskip

\section*{Introduction}

The present study was initially motivated by the broad topic of transport of particles in flows, which is of broad interest for fundamental scientific issues, industrial applications and natural systems. In nature, the formation of rain droplets in clouds, pyroclastic flows, dispersion of pollutants in the atmosphere or planet formation in accretion disks are just a few examples. In industrial applications, one can mention spray combustion in Diesel engines, pneumatic transport of granular media or mixing process.  Most of these situations are quite complex due to the turbulent nature of the carrier flow, which makes the coupling between the particle dynamics and the flow random and multi-scale. Further complexities arise, when particles can interact between each other (for instance if they are charged or magnetized).
\medskip

When it comes to the modeling of such particle laden flows, several situations must be distinguished depending on the nature of the particles. For sufficiently small particles, for which a point-wise approximation is relevant, the dynamics is well captured by the celebrated Maxey-Riley-Gatignol equation \cite{bib:maxey1983}. Depending on the size and the particle density, corrections to this equation (such as Fax\'en corrections) can be introduced to some extent~\cite{bib:calzavarini2009_JFM}. In the recent years, thanks to the development  of new technologies, in particular in the field of high speed digital imaging and particle tracking methods, the experimental study of turbulent transport has revolutionized our capacities to characterize the motion of particles, even in highly turbulent flows~\cite{bib:toschi2009_ARFM,bib:bourgoin2014_AGU}. Similar advances have also been achieved in numerical simulations, with the emergence of direct numerical simulations capable to resolve the full dynamics of the particles (beyond the point particle approximation) from first principles~\cite{bib:uhlmann2005_JCP}. Many of these experimental and numerical studies (see~\cite{bib:qureshi2007_PRL,bib:qureshi2008_EPJB,bib:volk2008_PhysicaD,bib:uhlmann2017_JFM} among others) have revealed the lack of a reliable theoretical framework to efficiently model the dynamics of finite size particles. Improving the usual models (or finding new modeling approaches) for the turbulent transport of particles remains crucial to account for subtle intricate effects such as wake effects on particle-flow interaction, the interplay between inertia and gravitational settling, the role of collective effects and clustering, etc.\cite{bib:Bourgoin2014}. 

In this broad context of inertial particle laden flows, the present work is a preliminary study in a long-term perspective aiming at addressing more particularly the role played by particle-particle interactions (which may become important in densely seeded situations) in the overall particle-turbulence interaction process. Such interactions are common in fluidized beds or in natural systems such as clouds, where water droplets tend to be charged.

In order to improve our understanding of the role played by the many possibly coupled phenomena above mentioned, it is wise to go step by step and consider first simpler situations in order to disentangle possible individual contributions of the different effects, before considering the fully coupled situation. In the present article, we therefore focus on some basic aspects of the role played by particle-particle interactions, considering the simplified situation where the flow is initially at rest and the particles are initially packed and also at rest. More precisely, we investigate the sedimentation of paramagnetic particles, subjected to gravity and to a tunable external homogeneous magnetic field, packed in a Hele-Shaw cell.  

The goal of the present study is to address the question whether such paramagnetic particles may be good candidates to address experimentally the impact of particle-particle interactions in flows laden with inertial particles (with density higher than that of the fluid). The interest of using paramagnetic particles relies on the fact that the amplitude of --dipole-dipole-- interactions between particles can be directly tuned by adjusting the amplitude of the applied magnetic field. The drawback is that the magnetic susceptibility of such particles is low (on the order of $10^{-5}$), and hence it is unclear whether the resulting magnetic forces between particles can reasonably compete with hydrodynamical forces. The chosen configuration for this preliminary study, with initially packed particles in a fluid at rest aims at considering first the most favorable situation regarding the possible impact of magnetic forces. This choice of such a configuration somehow moves the original motivation in the context of particle-laden flows, to the field of granular physics, where we investigate the decompaction of a granular system with particles with tunable interactions. Interestingly, beyond our longterm goal in the context of particle laden flows, this opens new perspectives for future dedicated studies with the same experimental setup to address interesting questions in the field of granular and deformable porous media.

Figure~\ref{fig:TempSeries} represents a time series of the investigated decompaction and sedimentation process. The situation is relatively canonical in the context of granular media research. Over the last decades, numerous studies have indeed investigated the sedimentation of such a granular pack in newtonian fluids \cite{Bagnold1954,bib:batchlor1972,bib:batchelor1986,bib:lange1997_EPJB,bib:voltz2001_PRE,bib:Vinningland2007a}. The question of the nature and the dynamics of interface instabilities in the early stage of the decompaction has received particular interest. Notably, the emergence of a fingering instability has been demonstrated, recalling the well-known Rayleigh-Taylor instability which classically develops at a horizontal interface between two fluids with different densities (the denser being on top)~\cite{bib:rayleigh1883,Taylor1950}. Several authors have investigated theoretically and experimentally the fingering instability of the particle pack / fluid interface in a Hele-Shaw cell \cite{bib:lange1997_EPJB,bib:voltz2001_PRE,bib:Vinningland2007a},
proposing  fluid interpretation \emph{\`a la} Rayleigh-Taylor. 
\medskip

The present study does not address the fingering instability (which will be considered in future studies) but focuses on characterizing the decompaction and fluidization rate of the particles pack when magnetic interactions are present. These interactions are driven here by the induced magnetization of the particles by the external field and are of the type \emph{induced dipole\--induced dipole}. As such, depending on the relative position and alignment of the particles the interactions can be either attractive or repulsive. It is therefore unclear, \emph{a priori}, whether they will contribute to an increased cohesion of the pack of particles or on the contrary they will accelerate its decompaction. Besides, considering the small value of the magnetic susceptibility of the paramagnetic particles considered, it is also \emph{a priori} unclear whether a measurable influence of the additionnal particle-particle magnetic interactions could be detected.

The article is organized as follows : in section~\ref{sec:I} we present the overall process of sedimentation of paramagnetic particles and the possible relevant dimensionless parameters of the problem; in section~\ref{sec:II} we describe the experimental setup ; section~\ref{sec:III} presents the main results, showing the impact of particle-particle interactions on the decompaction rate. The article ends with a brief discussion of the observed trends and possible perspectives of this work both in the context of particle laden flows and granular media.


\section{Sedimention of paramagnetic particles with an applied external magnetic field}\label{sec:I}

\begin{figure}[t] 
  
  \includegraphics[width=0.07\textwidth]{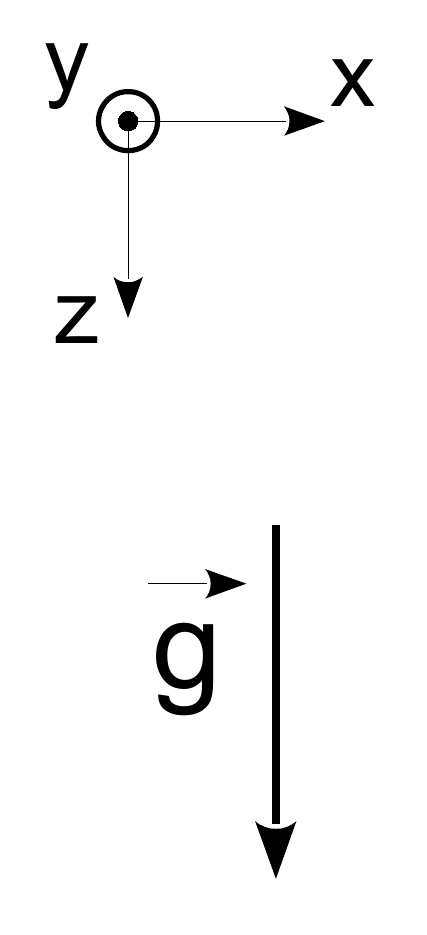}
  \centering
  \subfigure[t = 0s]{\includegraphics[width=0.18\textwidth]{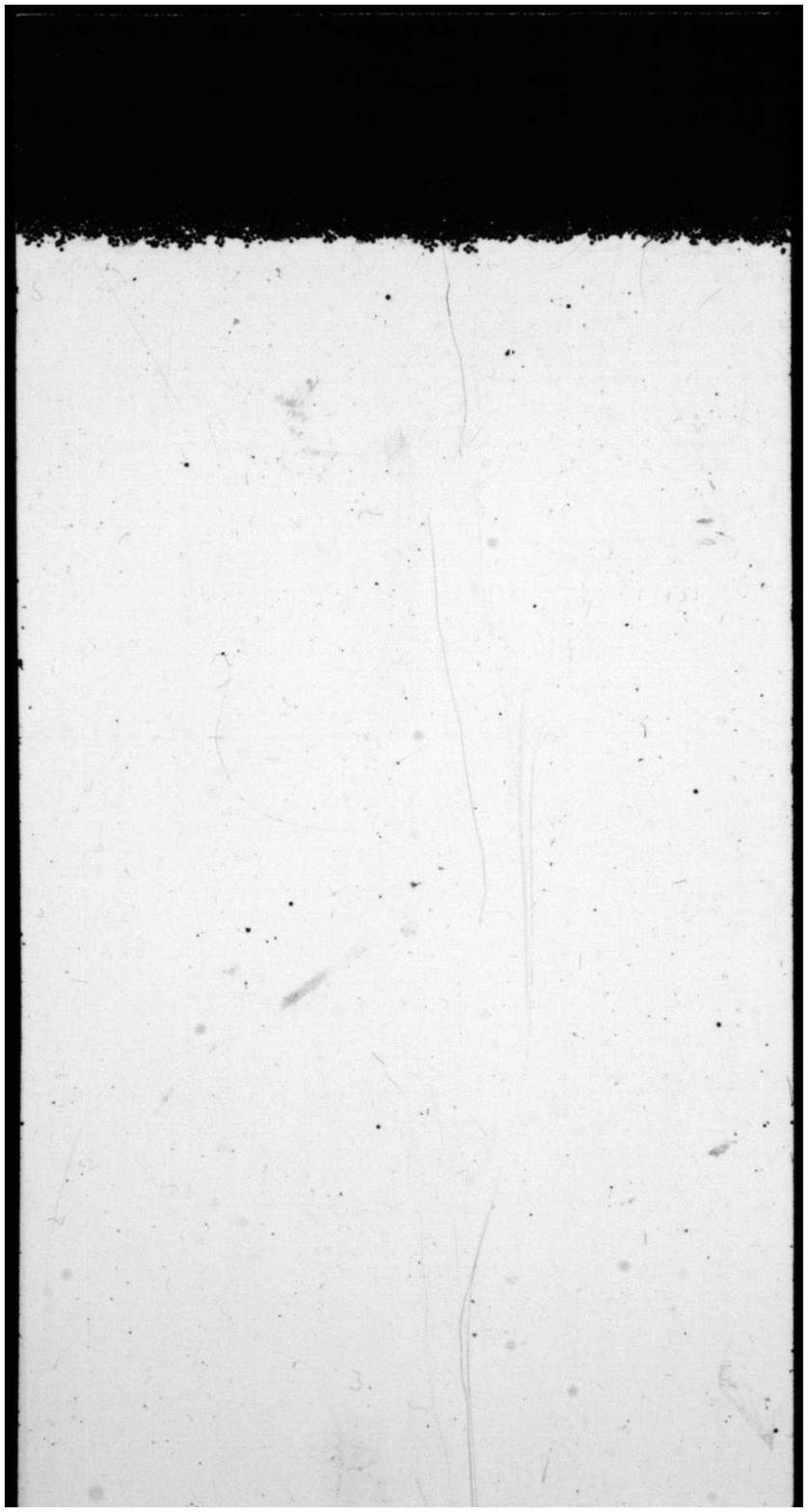}}
  \subfigure[t = 1s]{\includegraphics[width=0.18\textwidth]{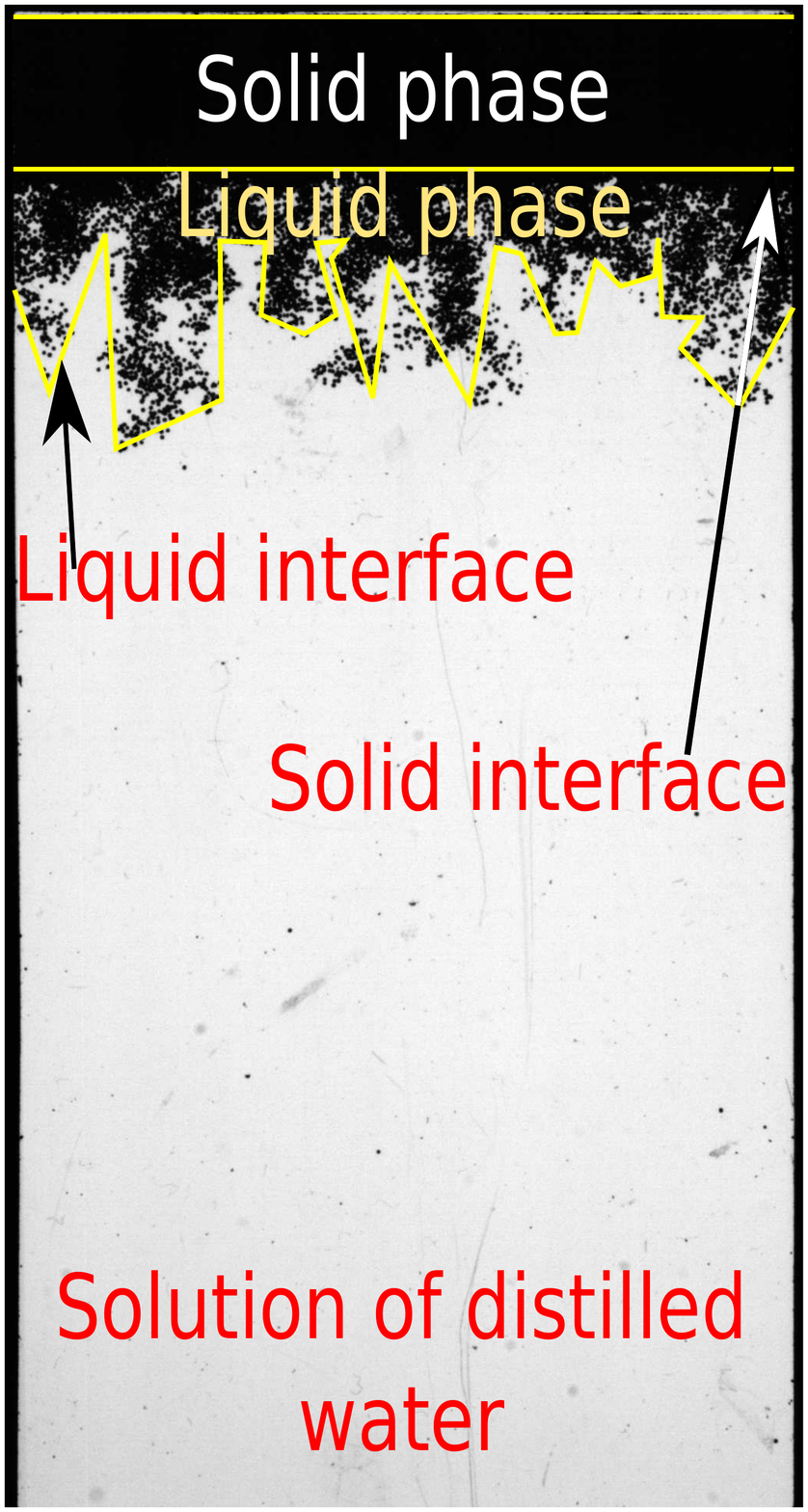}}
  \subfigure[t = 2s]{\includegraphics[width=0.18\textwidth]{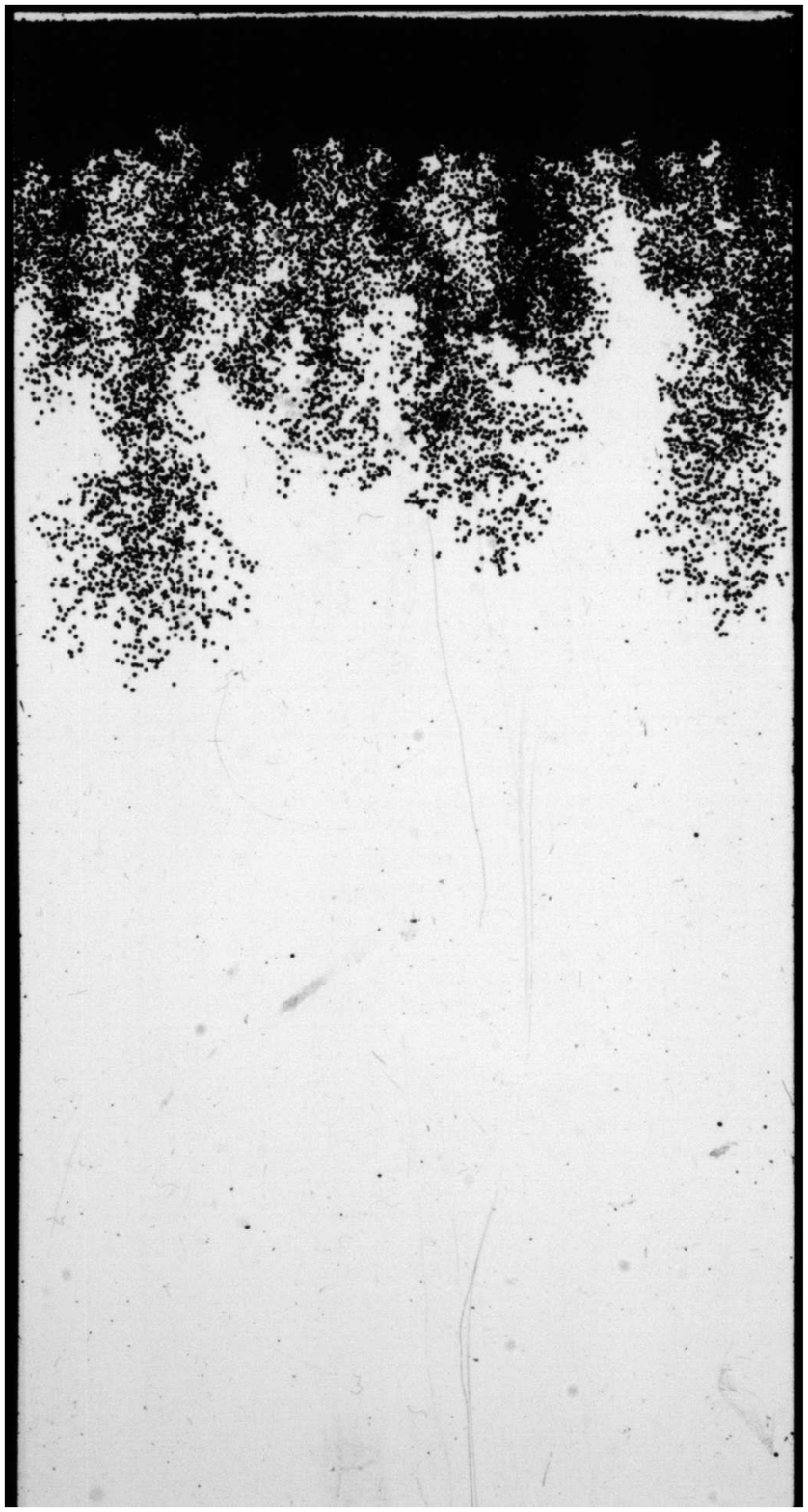}}
  \subfigure[t = 3s]{\includegraphics[width=0.18\textwidth]{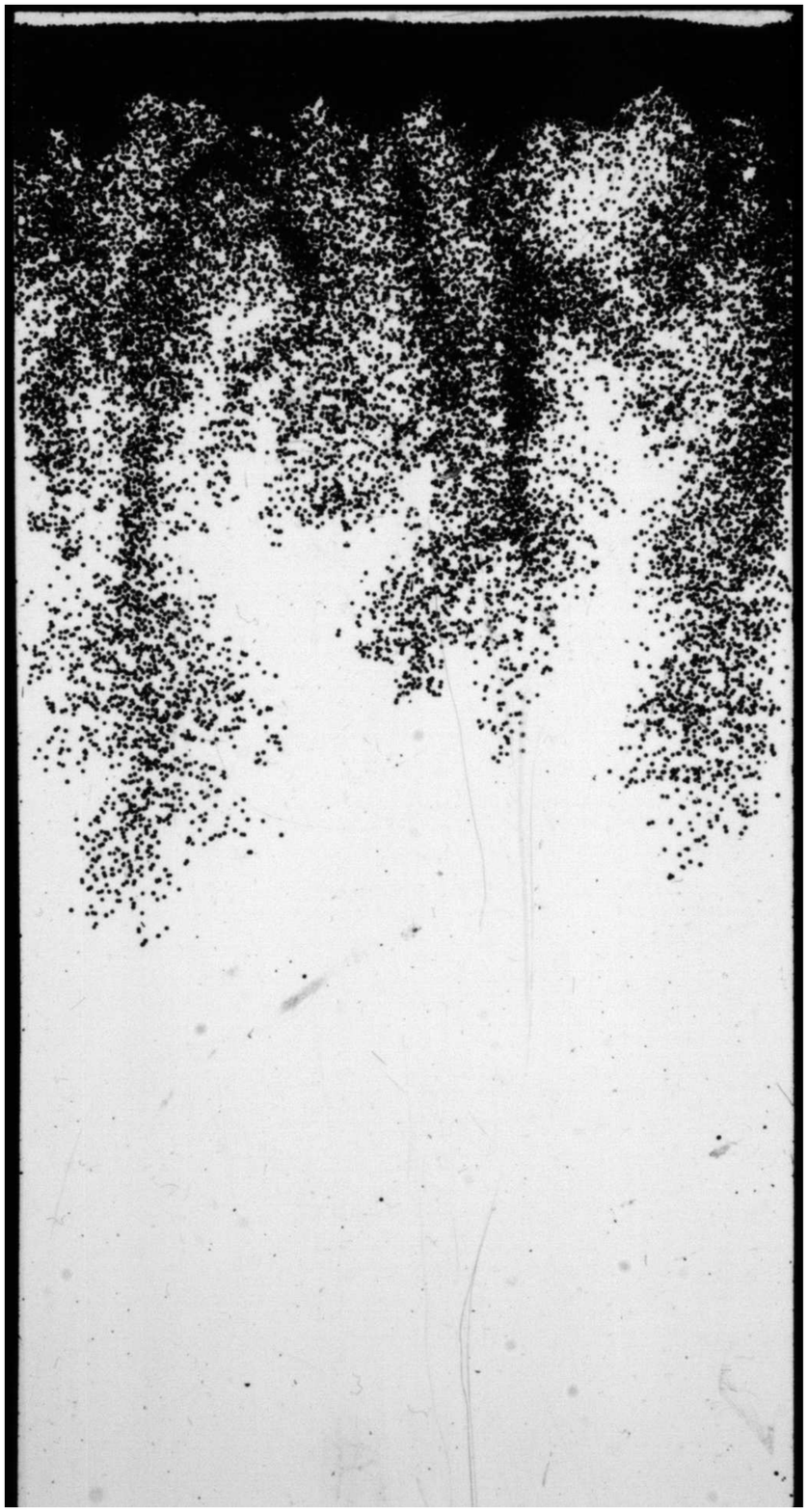}}
  \subfigure[t = 4s]{\includegraphics[width=0.18\textwidth]{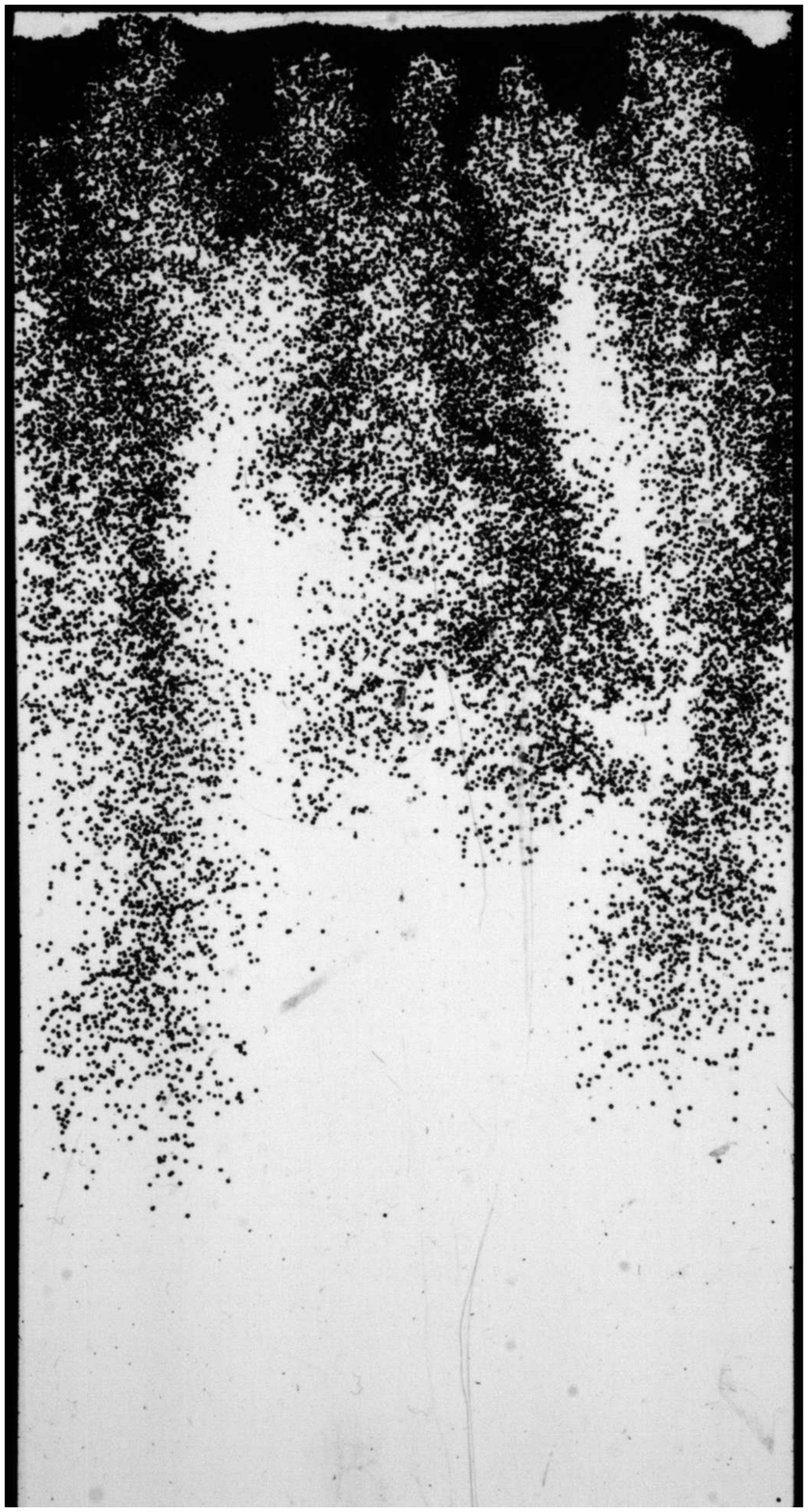}}
  \caption{{ Temporal series of images showing sedimentation subjected to gravity and an horizontal magnetic field in a steady flow with 250$\mu$m in diameter paramagnetic particles in a closed Hele-Shaw cell. The cell is 200 mm high, 50 mm wide and 1 mm depth.}}
  \label{fig:TempSeries}
\end{figure}

We briefly describe here the investigated decompaction and sedimentation process of dense particles in water as illustrated in Fig.~\ref{fig:TempSeries}. Particles are small (250~$\mu$m in diameter) monodisperse spherical beads with density $\rho=1.2$~g/cm$^3$. After the particle pack is prepared (see section~\ref{sec:II} for further details on the preparation step) the Hele-Shaw is rotated to bring the particle pack upwards. The time origin ($t=0$~s) is defined when the cell reaches the vertical position with the particle pack on top. On the first frame (Fig.~\ref{fig:TempSeries}a), we see the initial pack of particles, standing above water (note that the particles are fully permeated with water). The shape of the ``particle-bed/water'' interface is nearly flat and horizontal. On the next three frames (Fig.~\ref{fig:TempSeries}b-d), showing the subsequent decompaction and sedimentation of the particles, two main regions can be defined for the particle suspension: (i) in the upper part, particles remain packed and form a homogeneous and compact pack at rest, (ii) below the interface is disturbed giving rise to fingers of dispersed sedimenting particles. Following classical granular \emph{terminology}~\cite{bib:Jaeger1996}, we shall call \emph{solid} phase the upper compact region of particles and \emph{liquid} phase the lower sedimenting region. We therefore distinguish
two interfaces: (i) the first between the  \emph{solid} and  \emph{liquid} phases of particles, (ii) the second between the \emph{liquid} phase of particles and water. As the sedimentation process goes on, the volume of the \emph{solid} phase reduces (we shall refer to this as the \emph{decompaction} or the \emph{liquefaction} of the solid phase) and at the same time the area of the \emph{liquid} interface increases (the solid-liquid interface slowly moves up while the liquid-water interface develops larger fingers). In the last frame (Fig.~\ref{fig:TempSeries}e), the solid phase has totally \emph{melted} and disappeared. As the \emph{liquid} fingers of particles develop, the dynamics of the suspension becomes strongly  influenced by the induced recirculation flow of water in the cell and by lateral confinement effects due to the finite size  of the cell.

In the present work we focus on the dynamics of the initial decompaction process ($t \lesssim 3$~s) and do not address specific long-term finite size effects. We investigate the liquefaction rate of the \solid phase as well as the growth rate of the \liquid  phase, with the goal to highlight the  impact of particle-particle interactions on the global sedimentation behavior, when imposing a tunable homogeneous magnetic field $\vec{B}_0 = B_0\vec{z}$,  with $z$ the vertical direction such that gravity acceleration is $\vec{g} = g\vec{z}$.

In absence of any external magnetic field ($B_0 =0$~G), particles are only subjected to gravity $\vec{g}$ (and eventually hydrodynamic interactions). 

In presence of a homogeneous magnetic field ($B_0\small\neq$ 0~G), all particles acquire an identical induced magnetic moment $\vec{M}$ given by:
\begin{equation}\label{eq:M}
\vec{M} =  \dfrac{4}{3} \pi r^3 \dfrac{\chi_m}{\mu_0} \vec{B}_0,
\end{equation}
where $r$ is the radius of the particles, $\mu_0$ the vacuum permittivity and $\chi_m$ the magnetic susceptibility of the particles. In addition to gravity, each particle is then subjected to the magnetic force exerted by the magnetization of the other particles. The magnetic force $\vec{f}_m^i$ acting on particle $i$ can then be written as
\begin{equation}
\vec{f}^i_m =\left( \vec{M}_i \cdot\vec{\nabla}\right) \vec{B}_i,
\end{equation}
where $\vec{M}_i$ is the induced magnetization of particle $i$ and $\vec{B}_i=\vec{B}_0+\sum_{k \neq i} ^{N_p}\vec{b}_{ki}$ the total magnetic field at the position of particle $i$, including both the applied field $\vec{B}_0$ and the total induced magnetic field $\vec{b}_i=\sum_{k \neq i} ^{N_p}\vec{b}_{ki}$ from all  other particles than $i$. Note that as we consider here the case of a homogeneous applied magnetic field $\vec{B_0}$, there is no net force exerted by $\vec{B}_0$ on the magnetized particles and $f^i_m$ is simply given by 
\begin{equation}\label{eq:fi}
\vec{f}^i_m =\left(\vec{M_i} \cdot \vec{\nabla}\right)\vec{b}_i.
\end{equation}

We recall that the dipolar magnetic field $\vec{b}_{ki}$ generated by a particle $k$ with magnetization $\vec{M}_k$ at the position of particle $i$ is given by
\begin{equation}
\vec{b}_{ki} = \frac{\mu_0}{4\pi}\left(\frac{3\vec{r}_{ki}\left(\vec{M}_k\cdot\vec{r}_{ki}\right)}{|\vec{r}_{ki}|^5}-\frac{\vec{M}_k}{|\vec{r}_{ki}|^3}\right)
\end{equation}

\noindent where $\vec{r}_{ki}$ is the separation vector between particle $k$ and particle $i$. Considering the rapid decay of $\vec{b}_{ki}$ with $\left|\vec{r}_{ki}\right|$, contributions to $\vec{b}_i$ mostly come from the nearest neighbors, which in the initially compact pack are at a distance $\vec{r}_{ki}$ commensurate with the particle diameter $d=2r$ (although the orientation \--repulsive or attractive\-- of the force changes depending on the position of the considered neighbors around particle $i$). As a consequence, the typical expected order of magnitude for $\left|\vec{b}_i\right|$ is given by $\mu_0 M/4\pi r^3$, and considering Eqs.~(\ref{eq:M})~\&~(\ref{eq:fi}), the typical order of magnitude of the magnetic force is 
\begin{equation}
	{\cal O}\left(\left|\vec{f}^i_m\right|\right)\propto \frac{r^2 \chi_m^2B_0^2}{\mu_0}
\end{equation}

Based on these estimates, we can then define a dimensionless number $\Psi$ comparing the amplitude of magnetic particle-particle magnetic force and gravity (including effect of buoyancy) acting on each particle:

\begin{equation}\label{eq:Psi}
\Psi = \frac{\chi_m^2 B_0^2}{\mu_0 r \left|\rho_p-\rho_w\right| g}
\end{equation}

\noindent with $\rho_p$ and $\rho_w$ the density of the particles and of water respectively. We shall refer to $\Psi$ as the \emph{magnetic Bond} number. 

\section{Experimental Setup}\label{sec:II}

\begin{figure}[t]
\begin{minipage}[c]{.5\linewidth}
\centering
  \subfigure[Front view]{\includegraphics[width=0.45\textwidth]{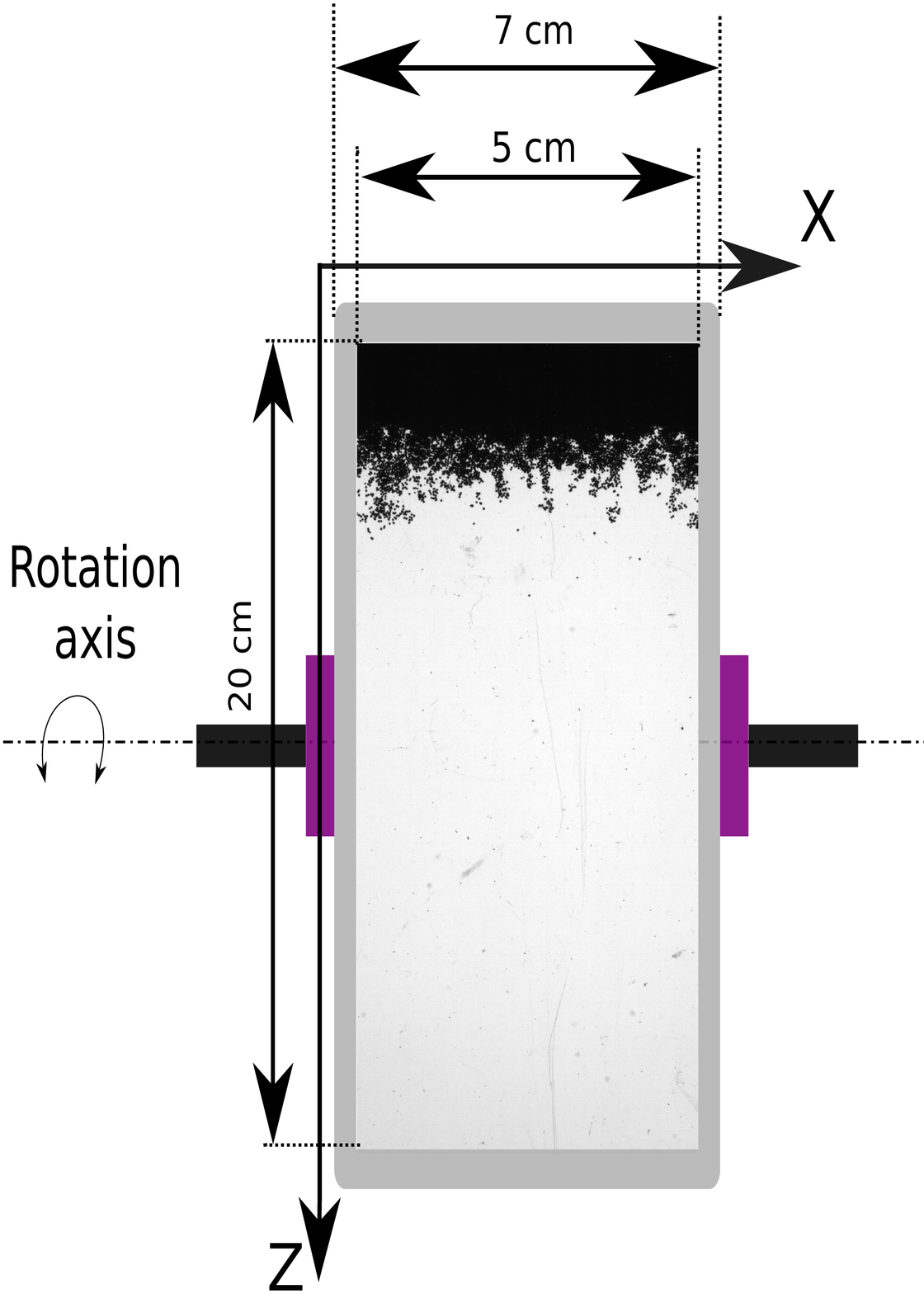}}\label{SchemFace}
  \subfigure[Side view]{\includegraphics[width=0.45\textwidth]{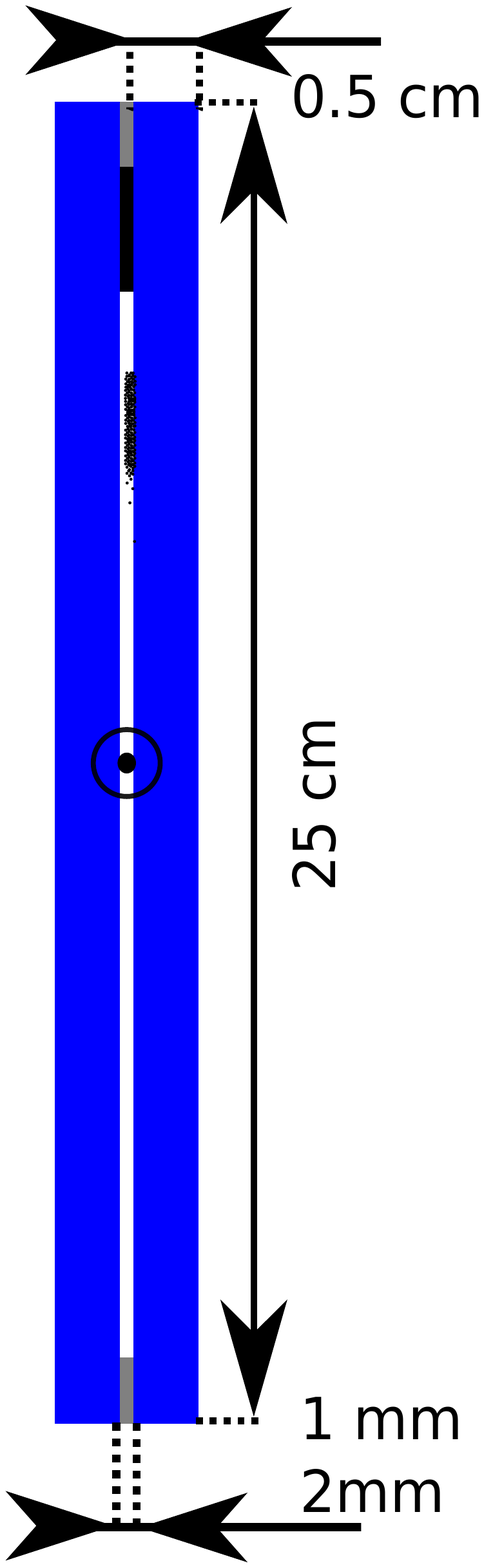}}\label{SchemCote}
  \caption{\textit{the Hele-Shaw cell.}}
  \label{Illust}
   \end{minipage} \hfill
   \begin{minipage}[c]{.5\linewidth}
	\centering
  \includegraphics[width=1\textwidth]{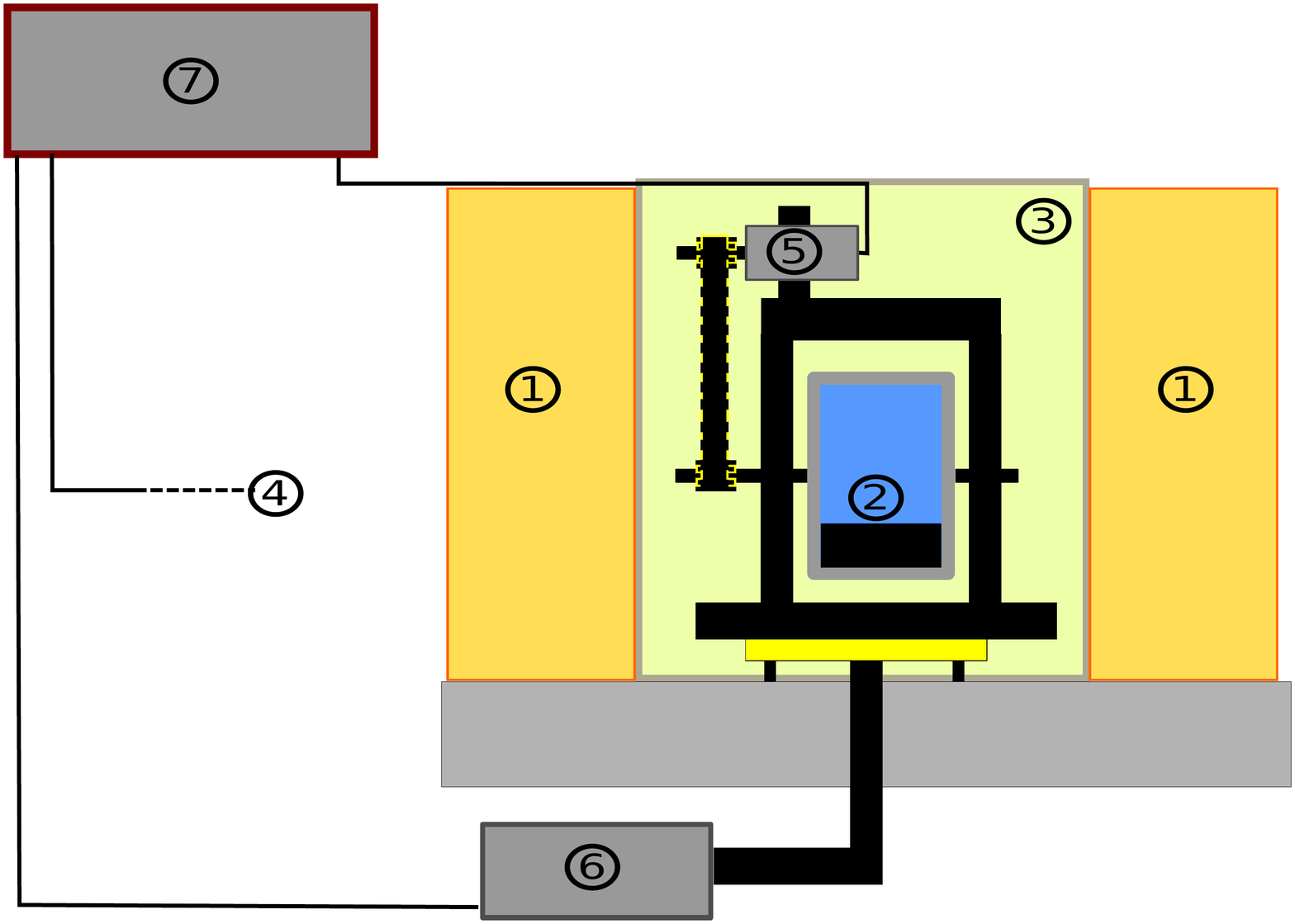}
  \caption{{Experimental setup : 1-coils, 2-Hele-Shaw cell, 3-LED panel, 4-High speed Camera, 5-stepper motor, 6-vibrator pot, 7-Raspberry-Pi.}}
  \label{ManipG}
   \end{minipage}
\end{figure}
   

Experiments were performed in a sealed Hele-Shaw cell (200~mm in height, 50~mm in width, with a gap of 1~mm), filled with distilled water and paramagnetic particles. A small amount of surfactant (Sodium dodecyl sulfate) is also added to stabilize the solution and to prevent aggregation of the particles. Fig.-\ref{Illust} shows a schematic of the cell and a typical recorded image of settling particles. The gap of the Hele-Shaw cell is fixed by a thin (1~mm in thickness) internal plastic frame clamped between two plexiglass plates. Two pivots screwed on the sides of the cell at mid-height allow to rotate the cell.
\\
As particles, we used 250~$\mu$m (in diameter) polythene (PE) micro-spheres, with density  $\rho_p=1.2$~g.cm$^{-3}$ doped with iron oxyde (FeO) in order to be paramagnetic (BKPMS particles from \textit{Cospheric LLC, Santa Barbara, CA USA}). We have determined their magnetic susceptibility $\chi_m$ from a SQUID magnetometer measurement on individual particles, carried at the CML (Center of Magnetometry of Lyon), leading to a value $\chi_m \simeq 4.2\cdot 10^{-3}$.


Two coils in Helmholtz configuration surround the cell in order to apply a homogeneous horizontal magnetic field $\vec{B}_0$, parallel to the x axis, aligned with the width of the cell~(Fig.~\ref{ManipG}). The coils are driven with a 10~kW power supply (\textit{Danfysik System 9100}). When fed by the maximum available current, a rapid increase of the coils temperature is observed, requiring an efficient cooling using a dedicated recirculation of cold water. The maximum amplitude of the applied magnetic field is $B_0^{max}\approx 0.12$~T.
\\
In order to carry a systematic investigation with good statistical convergence, for a given value of the applied magnetic field, of the order of 50 sedimentation experiments are repeated with the same initialization protocol. For a good reproducibility of experimental conditions, the setup is fully automatized, the rotation been ensured by a stepper motor controlled from a Raspberry Pi nano-computer with a pre-programmed cycle. Special care has been taken to properly condition the initial pack of particles in between rotations in order to have a particle-bed/water interface as flat and horizontal as possible before each rotation. This is achieved by applying to the celle several rapid oscillations and simultaneously vibrating the cell with an electrodynamic shaker. This optimal pack conditioning protocol is also automatized (and is part of the global rotation cycle controlled by the nano-computer) what warrants that the exact same protocol is applied for every experiment. {{Note that in spite of this care, the initial compaction of the particle bed may still present some variations from one realization to another. To limit any bias due to such variations of initial conditions, we remove from the ensemble of realizations those for which the initial height of the particle bed deviates from the average height by more than 10\%. This ensures considering realizations with almost identical initial compaction. As a consequence of this selection, of the order of 15\% of the realizations are dropped.}}

Measurements are based on high-speed backlight imaging of the cell, which is illuminated from behind by a LED panel. Recordings are performed using a high speed camera (\textit{Flare 2M360-CL from IO Industries}) operated with a resolution of 2048$\times$1088~pixels at a frame rate of 300~fps. The starting of the recordings is triggered by the controller nano-computer and is synchronized with the rotation cycle.
\medskip

\medskip

\medskip

\section{Experimental Results}\label{sec:III}

We present in this section results on the \emph{liquefaction} rate of the \emph{solid} phase and the simultaneous growth of the \emph{liquid} phase of the suspension in the early stage of the sedimentation, right after the cell reaches its vertical position. To quantitatively address these questions we use image processing tools to detect the \emph{solid-liquid} and \emph{liquid}-water interfaces.
\newline
\subsection{Edge detection \& area calculation}
\medskip

\begin{figure}[t]
  \centering
  \subfigure[Raw image]{\includegraphics[width=0.15\textwidth]{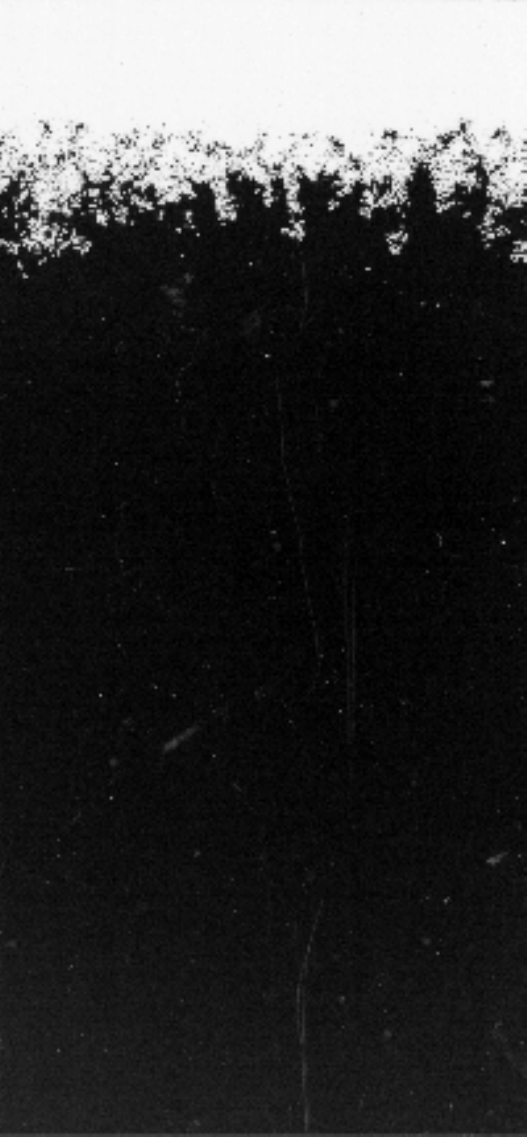}}
  \subfigure[Low-pass filtered image]{\includegraphics[width=0.15\textwidth]{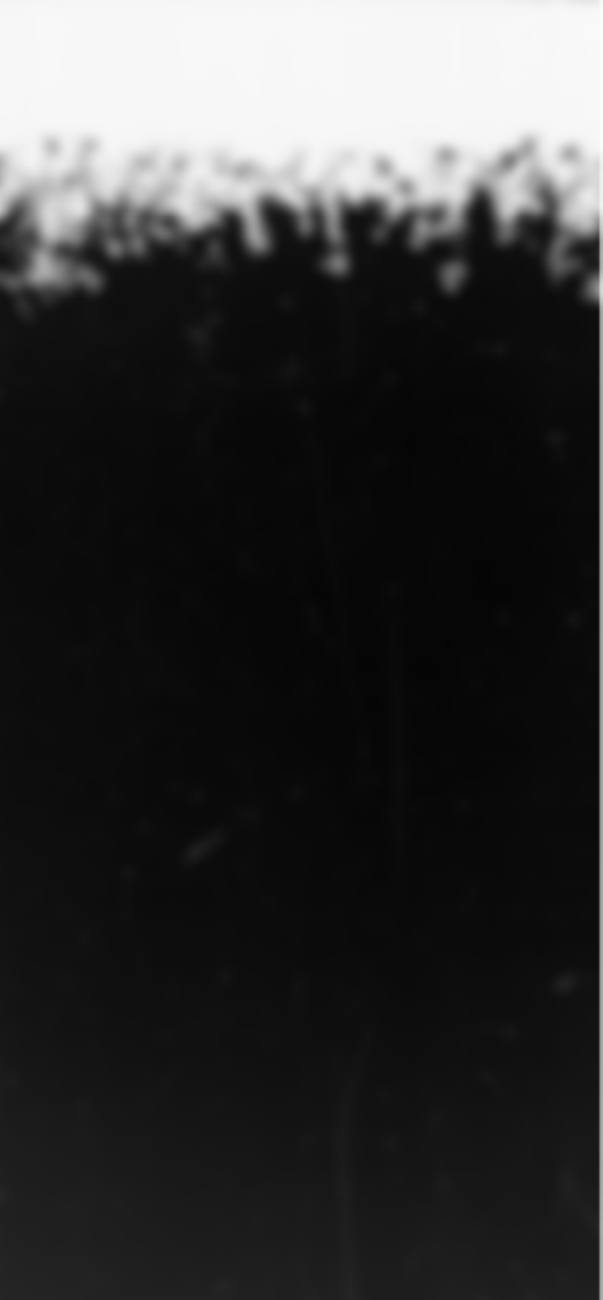}}
  \subfigure[Egdes detection]{\includegraphics[width=0.15\textwidth]{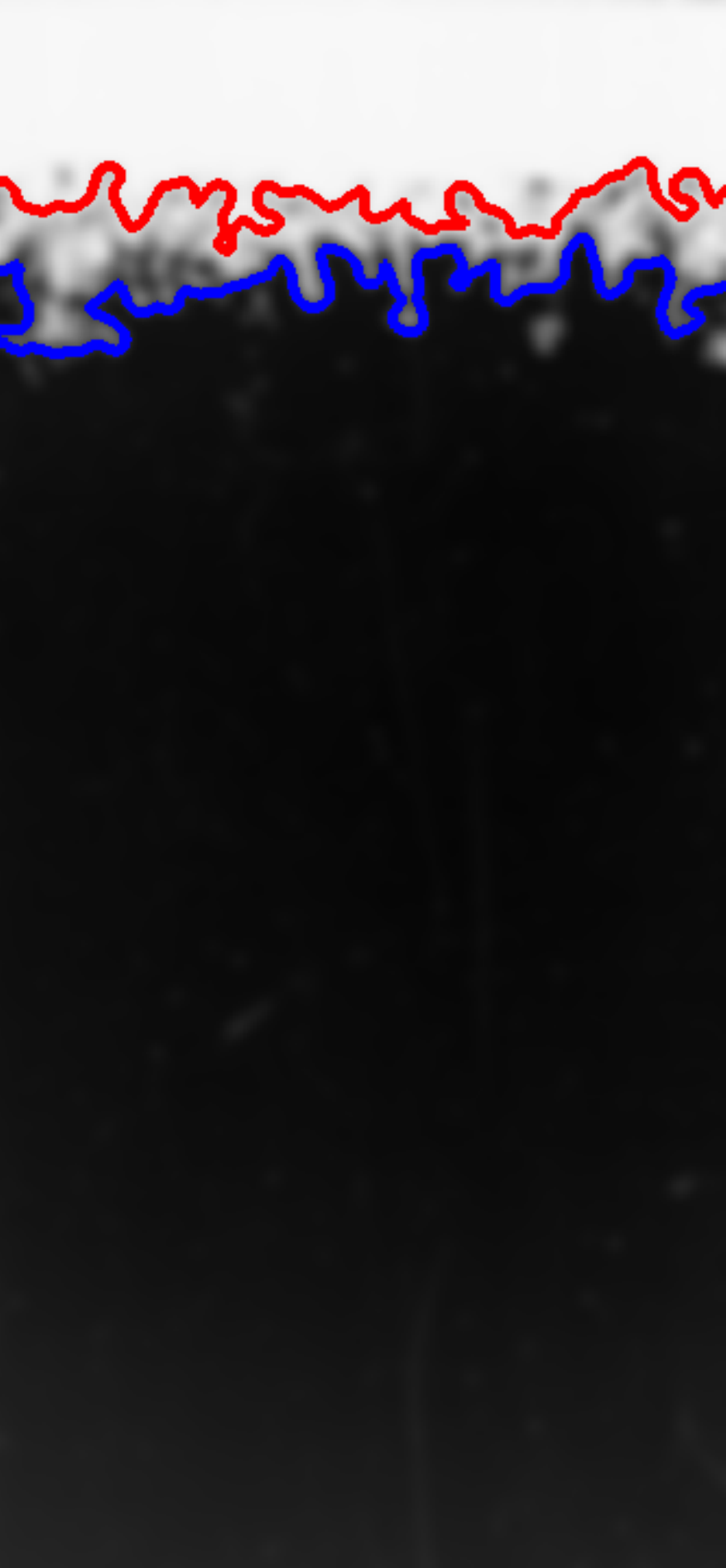}}
      \subfigure[Ternarized image]{\includegraphics[width=0.13\textwidth]{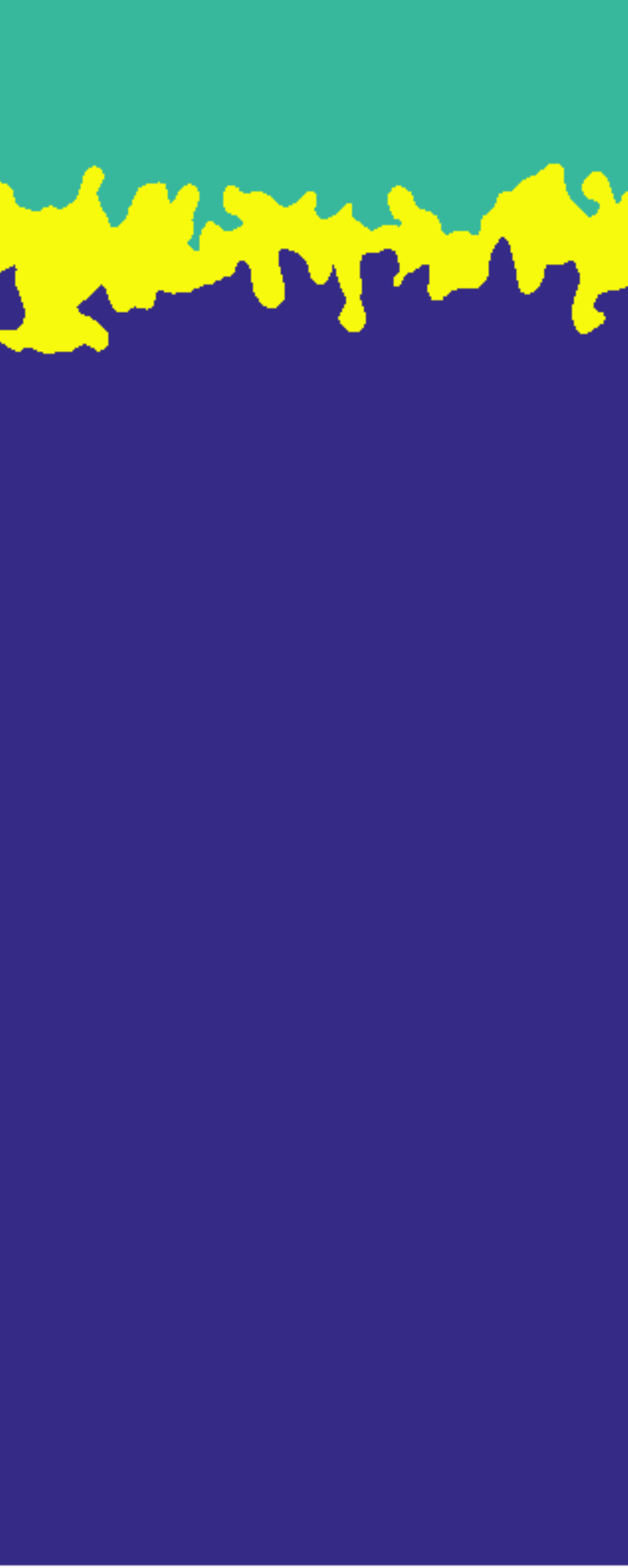}}
  \caption{{Successive steps of image processing to determine the \emph{solid} phase (upper green area in fig.(d)) and \emph{liquid} phase (middle yellow area in fig.(d)) of particles starting from the raw image (fig.\ref{fig:imanalysis}a).}}
  \label{fig:imanalysis}
\end{figure}

The successive image processing steps are illustrated in Fig.~\ref{fig:imanalysis} (note that, compared to previous figures, negative images are shown, particles appear now as bright).

Fig.~\ref{fig:imanalysis}(a) shows a raw image, where the \emph{solid} phase appears as a continuous bright region on the top whereas the \emph{liquid} phase is characterized by some fragmentation of the intensity pattern below. We first apply a Gaussian low-pass filter in order to blur the image in the liquid phase: as a result (i) nearby particles in the fragmented region become connected and (ii) the locally smoothed intensity in the \emph{liquid} region decreases as shown in Fig.~\ref{fig:imanalysis}(b). We use a Gaussian convolution kernel with a width of 1.5 particle diameter in order to reconnect the fragmented areas up to this scale. Few blobs of particles separated by more than 1.5~$d$ can still be present; those will be considered as having already left the \emph{liquid} phase and are excluded from the detection. Note that the Gaussian filtering has no major effect on the \emph{solid} phase region of the image which has constant and homogeneous intensity. Then, two successive thresholdings of the image are operated. The first uses a relatively low threshold chosen in order to keep all the particles (non-dark pixels) in both the \emph{solid} and \emph{liquid} phases. An edge detection algorithm is then run on the binarized image to determine the \emph{liquid}-water interface (blue line in Fig.~\ref{fig:imanalysis}(c) ). The second thresholding uses a higher value of the threshold, chosen to keep only the bright pixels in the continuous region corresponding to the \emph{solid} phase. Another edge detection then retrieves the \emph{solid-liquid} interface (solid red line in Fig.~\ref{fig:imanalysis}(c)). The two different binarized images can also be combined to create a ternarized image showing the \emph{solid} phase (upper green region in Fig.~\ref{fig:imanalysis}(d)), the \emph{liquid} phase (middle yellow region in Fig.~\ref{fig:imanalysis}(d)) and water (lower blue region Fig.~\ref{fig:imanalysis}(d)).
\medskip

This image processing sequence is performed for all frames of each recording, in order to monitor the time evolution of the areas $A_s(t)$ and $A_l(t)$ of the \emph{solid} and \emph{liquid} phases respectively.
\\
In the sequel, for a given experiment (\emph{i.e.} a given value of the applied magnetic field amplitude $B_0$) we investigate the time evolution of the \emph{solid} and \emph{liquid} phase areas ensemble-averaged over the full set of realizations:

\begin{equation}
\left<A_{s,l}\right>(t) = \dfrac{1}{N_{movie}}\sum^{N_{movie}}_{k=1} A_{s,l}^{k}(t-t_{0,k}),
\end{equation}

\noindent with $N_{movie}$ the total number of movies (realizations) recorded for each experiment ($N_{movie}\gtrsim 40$ for each value of applied magnetic field amplitude $B_0$). Note that $k$ denotes the frame number corresponding to time $t$ for each movie, where the reference time $t_{0,k}$ to the instant where the cell reaches its vertical orientation after each rotation. For the sake of clarity, we shall omit from now the average brackets when presenting the results and simply refer to $A_{s,l}(t)$ for the ensemble-averaged areas of the \emph{solid} and \emph{liquid} phases.
 
\subsection{Liquefaction dynamics and particle concentration}
\subsubsection{Evolution of \emph{solid} and \emph{liquid} phases}
\begin{figure}[t]
  \centering
    \subfigure[Mean area versus time]{\includegraphics[width=0.45\textwidth]{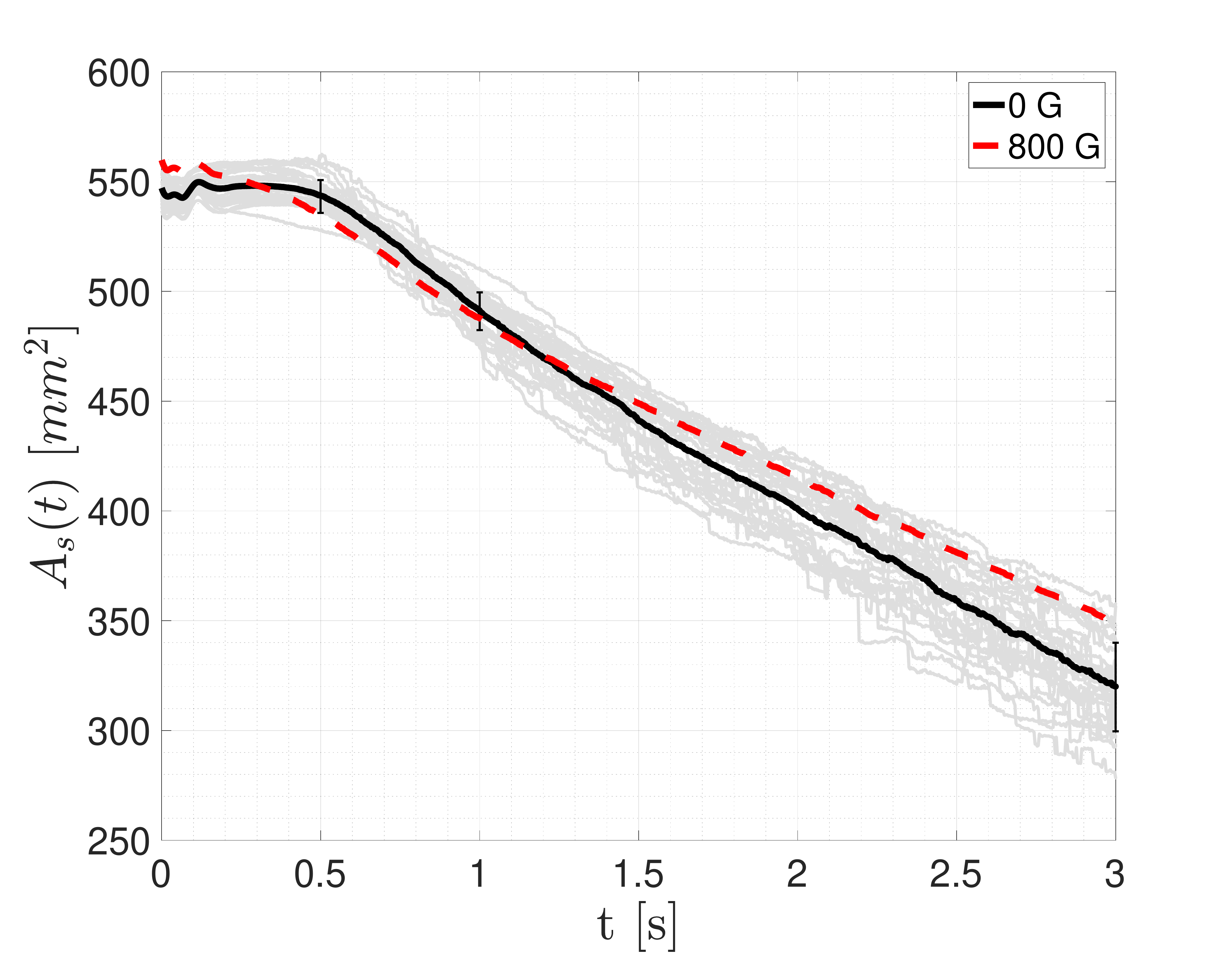}}
  \subfigure[Mean velocity versus time]{\includegraphics[width=0.45\textwidth]{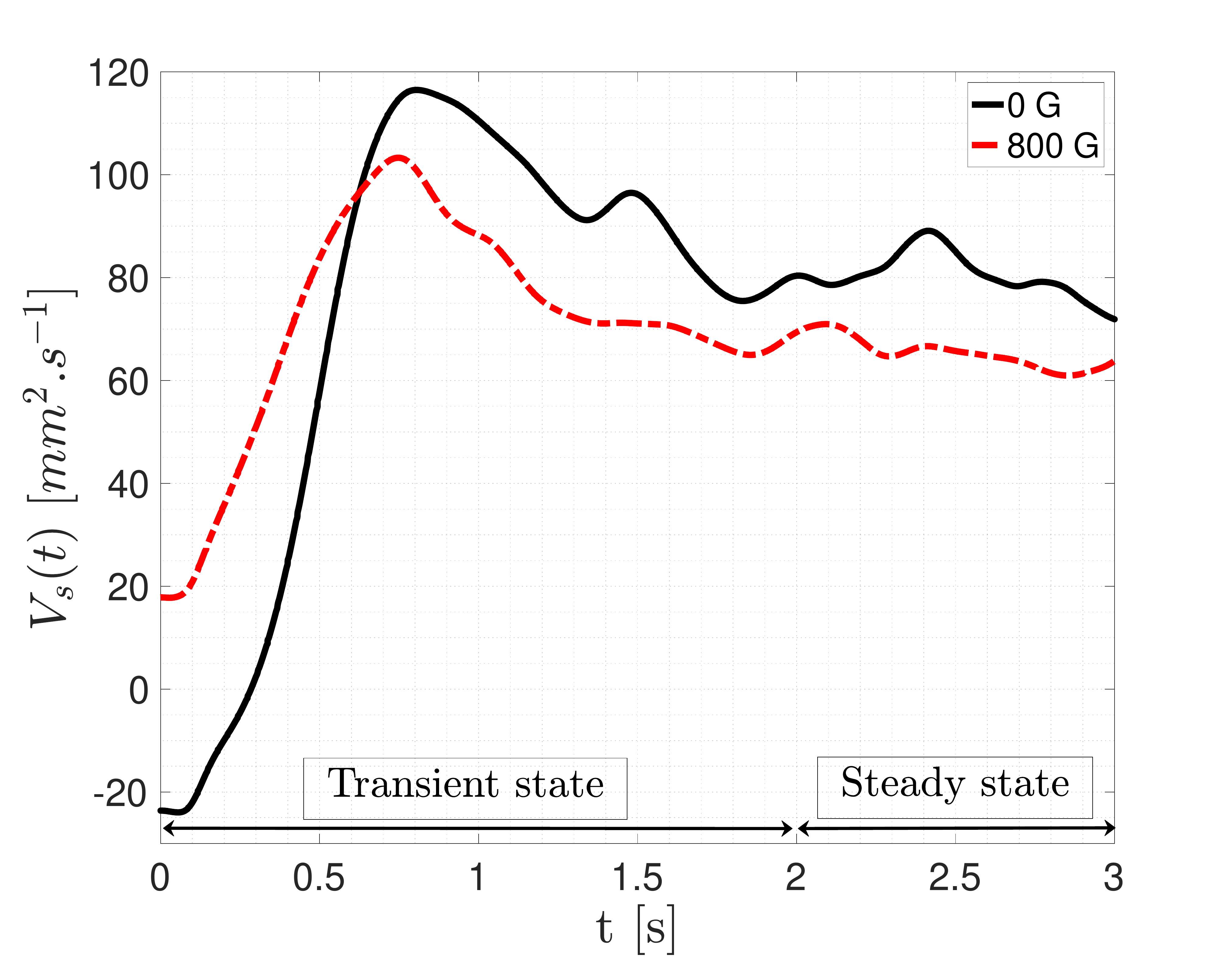}}
  \caption{{(a) Averaged time evolution of the solid area without any applied magnetic field (solid black line) and with an applied field with amplitude $B_0 = 800$~G (dashed red line, corresponding to a magnetic Bond number $\Psi \simeq 0.26$). To illustrate the averaging procedure, the light gray lines represent the individual time evolution of the solid area for 50 independent realizations for the case  without any applied magnetic filed. The solid black line correspond to the average of the light gray lines. The bars at 0.5~s, 1~s and 3~s indicate the standard deviation among the 50 realizations at each of these instants. (b) Time evolution of the \emph{liquefaction} velocity for $B_0=$~0~G (solid black line) ; $B_0=$~800~G (dashed red line)}.}
  \label{AreaVel}
\end{figure}

Fig.~\ref{AreaVel}(a) shows the time evolution of the \emph{solid} phase area $A_s(t)$ for the case without any applied magnetic field ($B_0=0$~G) and for the case with $B_0=800$~G ($\Psi\simeq 0.26$). A common pattern can be seen where after a short period (of the order of 300~ms) with almost no apparent evolution, a rapid decrease of $A_s(t)$ occurs followed by a slower linear decrease. A first visible impact of the magnetic field appears in the long term behavior, where we see that the \emph{solid} phase \emph{liquefies} slower in the presence of an applied magnetic field. 
At $t=3$~s we observe indeed that the remaining area of \emph{solid} phase is the largest for $B_0=800$~G and the smallest when no magnetic field is applied.

In order to further characterise the influence of the magnetic field on the acceleration/deceleration of \emph{liquefaction} we introduce the \emph{liquefaction rate}:
\begin{equation}
V_{s}(t) = -\dfrac{d A_s (t)}{dt}. 
\end{equation}

\noindent Note that $V_s$ has units of m$^2$s$^{-1}$.
Fig.~\ref{AreaVel}(b) shows the time evolution of the \emph{liquefaction} rate. Two regimes are clearly visible: after a short transient with a rapid increase of the \emph{liquefaction} rate followed by a local maximum, the a constant asymptotic terminal value $V_s^\infty$ is reached. It can be noted that in spite of the large number of recorded movies for each experiment, while the raw curves for $A_s(t)$ appear relatively smooth and well converged, their derivative still exhibit some visible noise. We therefore estimate $V_s^\infty$ by averaging $V_s(t)$ typically on the time interval $t\in[2~\textrm{s}-3~\textrm{s}]$. Equivalently, we can estimate the terminal \emph{liquefaction} rate from a linear fit of the long term trend of $A_s(t)$.


During the initial transient phase we can notice the presence of the small peak at very short time (around $t\approx 150$~ms), reminiscent of a small damped oscillation of the cell when it stops at its final vertical position just after the rotation. In the subsequent increase of the \emph{liquefaction} rate a systematic trend can be observed where the larger the magnetic field, the larger the \emph{liquefaction} rate. The mild overshoot preceding the terminal regime occurs at a similar time $t\approx 0.7$~s regardless of the amplitude of the magnetic field, although it seems slightly more peaked as the amplitude of the magnetic field increases. 

\begin{figure}[h!]
  \centering
     {\includegraphics[width=0.49\textwidth]{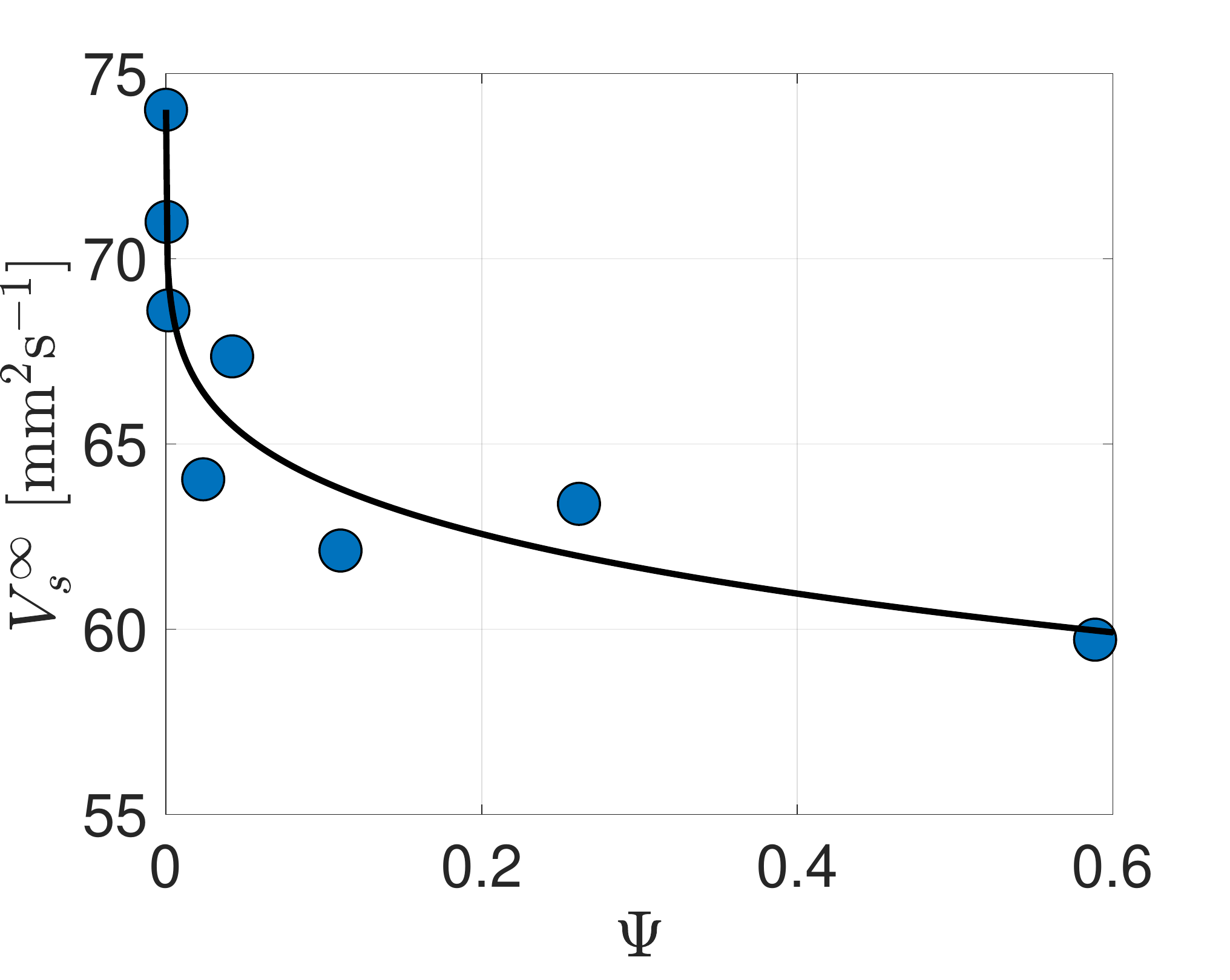} \label{VMax}}
  {\includegraphics[width=0.49\textwidth]{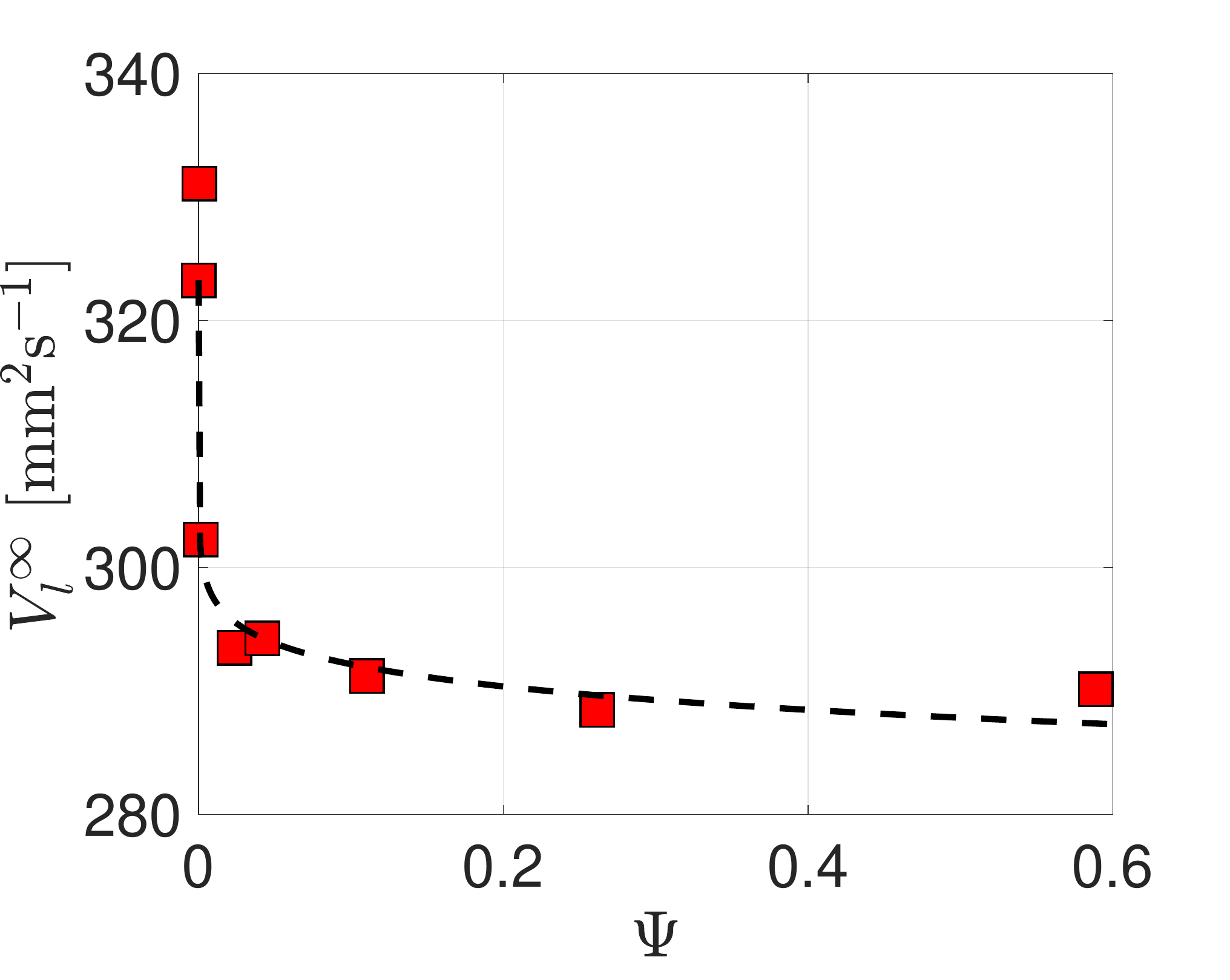}\label{Vmean}}
  \caption{{(a) Terminal \emph{liquefaction} rate of the \emph{solid} phase as a function of the amplitude of the magnetic Bond number, representing the amplitude of the applied magnetic field. The dot-dashed line represent the power law trend in eq.~\ref{eq:powSolidLiquid} with $a_s = 0.54 \pm 0.2$ and $b_s = 0.19\pm 0.08$. (b) Terminal growth rate of the \emph{liquid} phase as a function of the magnetic Bond number. The dot-dashed line represent the power law trend in eq.~\ref{eq:powSolidLiquid} with $a_l = 0.17 \pm 0.05$ and $b_s = 0.08\pm 0.04$. }}
  \label{VsVlmean}
\end{figure}

Concerning the terminal \emph{liquefaction} regime, Fig.~\ref{VsVlmean}(a) shows the \emph{liquefaction} average terminal rate $V_s^\infty$ as a function of the magnetic Bond number. In agreement with the qualitative observed trend of $A_s(t)$, we see that, in spite of some scatter, the terminal \emph{liquefaction} rate of the \emph{solid} phase globally decreases when the magnetic field increases. 

A similar analysis has been performed for the growth of the \emph{liquid} phase, defining the \emph{liquid} growth rate as:
\begin{equation}
V_{l}(t) = \dfrac{d A_l (t)}{dt}. 
\end{equation}

Fig.~\ref{VsVlmean}(b) shows the \emph{liquid} average terminal growth rate as a function of the magnetic Bond number. Trends similar to those of the \emph{liquefaction} of the \emph{solid} phase are observed, with a global decrease of the \emph{liquid} growth rate as the magnetic field increases. 

\begin{figure}[t]
    \centering
    \includegraphics[width=.65\textwidth]{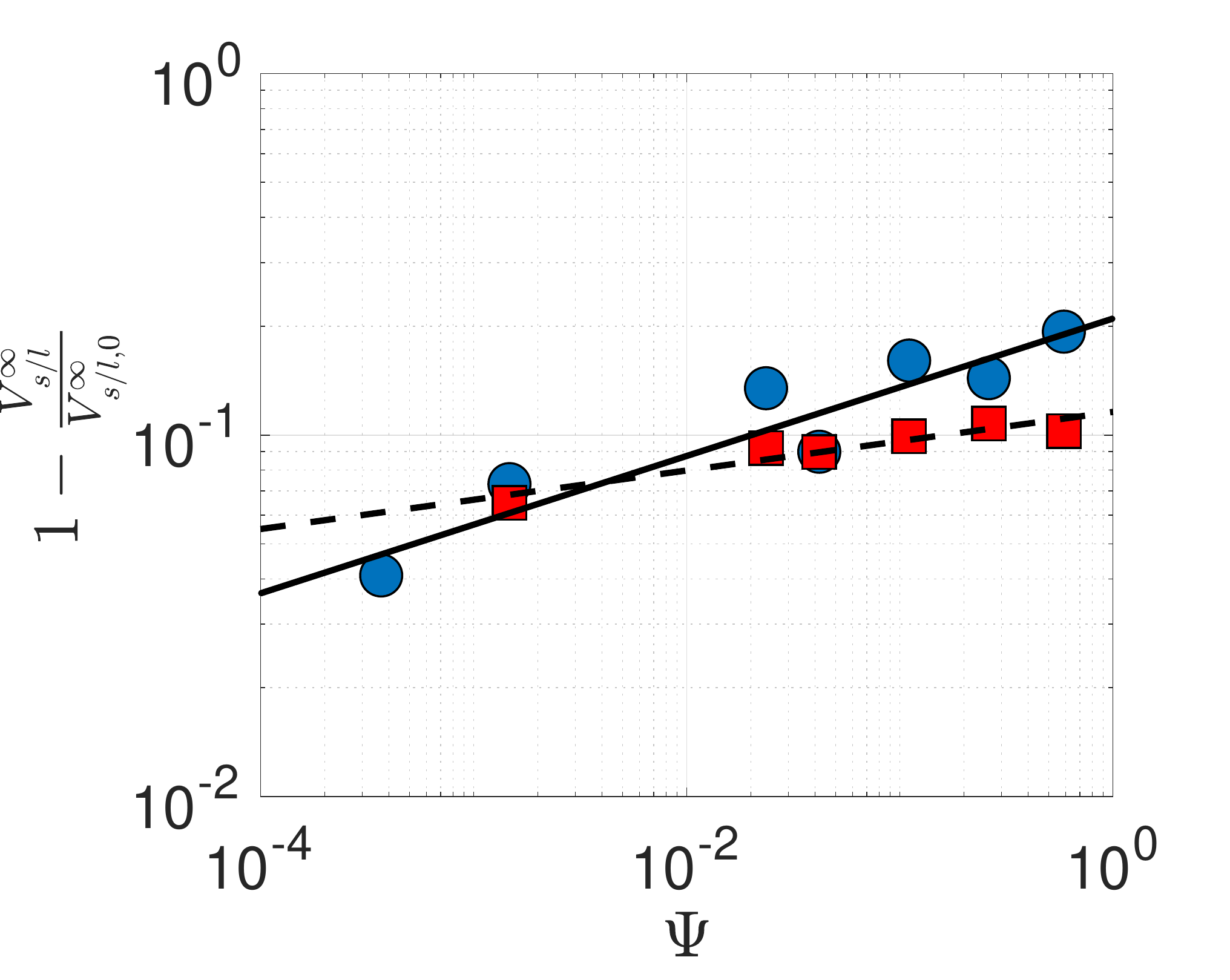}
    \caption{Relative liquefaction (for the \emph{solid} phase, blue circles ($\circ$)) and growth rate (for the \emph{liquid} phase, red squares($\square$)) as a function of the magnetic Bond number. Dot-dashed and solid lines correspond to the power-law fit~(\ref{eq:powSolidLiquid}).}
    \label{fig:VsVladim}
\end{figure}

The observed trends for the \emph{solid} and the \emph{liquid} phases show consistently that as the magnetic field amplitude increases, the \emph{solid} phase is stabilized (in the sense that it \emph{liquefies} slower) while the \emph{liquid} phase is less supplied with particles and hence it also grows slower. To further characterize the trends of the terminal liquefaction rate of the \emph{solid} phase and the terminal growth rate of the \emph{liquid} phase, we introduce the relative terminal rates defined respectively as $1-V_s^\infty/V_{s,0}^\infty$ and $1-V_l^\infty/V_{l,0}^\infty$ where the subscript $0$ indicates the values when no magnetic field is applied ($B_0=0$). Fig.~\ref{fig:VsVladim} shows a log-log representation of the relative terminal rates as a function of the magnetic Bond number $\Psi$. For both the \emph{solid} and the \emph{liquid} phase, the trends are reasonably fitted by a power law such that 
\begin{equation}\label{eq:powSolidLiquid}
    1-\frac{V_{s/l}^\infty}{V_{s/l,0}^\infty} = a_{s/l} \Psi ^{b_{s/l}},
\end{equation}

\noindent with 

\begin{equation}\label{eq:solid}
    a_s = 0.21 \pm 0.08 \;\; , \;\; b_s = 0.19\pm 0.08
\end{equation}

\noindent and

\begin{equation}\label{eq:liquid}
        a_l = 0.12 \pm 0.01 \;\; , \;\; b_s = 8.1 \cdot 10^{-2} \pm 0.04.
\end{equation}

These trends reveal that the influence of the magnetic field is significantly stronger in hindering the liquefaction of the \emph{solid} phase, than in hindering the growth rate of the \emph{liquid} phase. This could be expected considering that in the more dilute \emph{liquid} phase the inter-particle distance is larger than in the compact \emph{solid} phase where particles are in contact with each other. The induced magnetic particle-particle interactions are then reduced in the \emph{liquid} phase which is indeed found to be marginally affected by the applied magnetic field. An interesting question then concerns the evolution of the particle concentration (and hence of the inter-particle distance) in the \emph{liquid} phase. This particle concentration results from the balance between the particles being supplied from the \emph{liquefaction} of the \emph{solid} phase and of the growth with time of the \emph{liquid} area $A_l(t)$ as the \emph{liquid} phase develops. We investigate this balance in the next sub-section.



\subsubsection{Particle concentrations in the \emph{solid} and \emph{liquid} phases}
\begin{figure}[h!]
  \centering
     \includegraphics[width=0.45\textwidth]{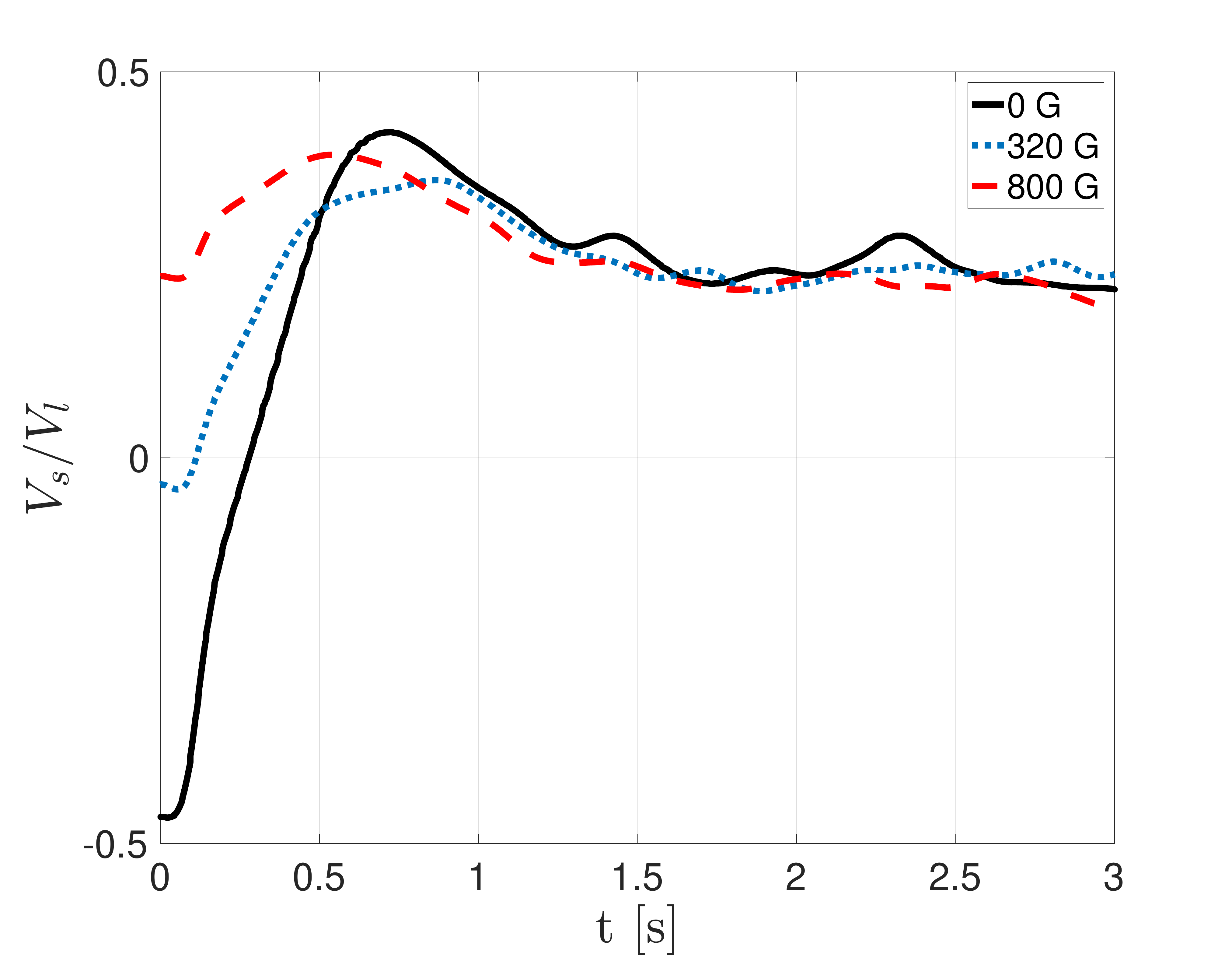}
  \caption{Evolution with time of the \emph{solid} to \emph{liquid} rate ratio. In the terminal steady state, the \emph{solid} to \emph{liquid} rate ratio reaches an asymptotic value of the order of 0.25, independent of the magnetic field amplitude, indicating that the \emph{liquid} phase is about four times more dilute than the \emph{solid} phase.}
  \label{Rapp}
\end{figure}

Let $\rho_l$ and $\rho_s$ be the particle concentration (number density) in the \emph{liquid} and the \emph{solid} phase respectively. The particle concentration $\rho_s$ in the \emph{solid} phase is assumed to be constant and fixed by the compacity of the initial pack. 
Initially, the \emph{liquid} phase is empty and emerges when the \emph{solid} phase starts \emph{liquefying}. The particle concentration in the \emph{liquid} phase is therefore expected to have a non-trivial time evolution. 

Keeping in mind that the cell is sealed, and hence the total number $N_p$ of particles remains constant, the density in both phases can be directly related to the rates of \emph{liquefaction} of the \emph{solid} phase $V_s$ and the rate $V_l$ at which the \emph{liquid} phase grows. Assuming that all particles remain either in the \emph{liquid} or the \emph{solid} phase (this approximation neglects the few particle blobs eventually detaching from the \emph{liquid} phase) and that the concentration within each phase is relatively homogeneous, we can indeed write

\begin{equation}\label{eq:Np}
N_p=\rho_l(t) A_l(t) + \rho_s(t)A_s(t).
\end{equation}

The conservation of the total number of particles then imposes $\textrm{d}N_p/\textrm{d}t=0$ which, considering Eq.~\ref{eq:Np} and assuming that $\rho_s$ remains constant, leads to

\begin{equation}\label{eq:Gamma}
\dfrac{V_s(t)}{V_l(t)}=\Gamma_{ls}(t)+ \dfrac{d \Gamma_{ls}(t)}{d t} \dfrac{A_l(t)}{V_l(t)},~~~~~\textrm{with}~~~~~\Gamma_{ls}(t)=\dfrac{\rho_l(t)}{\rho_s(t)}.
\end{equation}

This relation directly links the \emph{liquid} to \emph{solid} concentration ratio $\Gamma_{ls}$ to the corresponding areas and growth rates and in particular to the \emph{solid} to \emph{liquid} rate ratio $V_s(t)/V_l(t)$, which is represented in Fig.~\ref{Rapp}. In principle, Eq.~\ref{eq:Gamma} is a simple differential equation for $\Gamma_{ls}$ with empirically known time dependent coefficients ($V_s(t)/V_l(t)$ and $A_l(t)/V_l(t)$, which are directly accessible from the measurements), from which we could derive the time evolution of $\Gamma_{ls}$. We focus here though on the long term asymptotic behavior, where the ratio $V_s(t)/V_l(t)$ is constant, and found to be of the order of $V_s^\infty/V_f^\infty\approx 0.25$ independently of the amplitude of the applied magnetic field (see Fig.~\ref{Rapp}). In this stationary regime, the solution of Eq.~\ref{eq:Gamma} is $\Gamma^\infty_{ls}=V_s^\infty/V_f^\infty$, hence $\Gamma^\infty_{ls}\approx 0.25$ independently of the amplitude of the applied magnetic field. This means that the \emph{liquid} phase is about four times more dilute than the \emph{solid} phase, and that the particle average concentration in the \emph{liquid} phase is not affected by the presence of the magnetic field.
\medskip

\section{Discussion and Conclusion}
Our experiments reveal that, even if the magnetic Bond numbers investigated in the present study remains relatively low (not exceeding 0.6), measurable effects are still observed regarding the impact of an applied magnetic field on the decompaction rate under gravity of a pack of paramagnetic particles. Overall the influence of an increasing magnetic field on the decompaction of the \emph{solid} pack exhibit several interesting features :

\begin{itemize}
\item In the initial phase, the \emph{liquefaction} is accelerated in presence of the magnetic field;

\item The terminal \emph{liquefaction} rate is on the contrary reduced by the magnetic field and so is the growth rate of the \emph{liquid} phase, although the impact of the magnetic field on the \emph{liquid} phase remains marginal;

\item The terminal concentration of the \emph{liquid} phase is about a quarter that of the \emph{solid} phase and does not depend on the applied magnetic field.
\end{itemize}

Some features of these trends can be qualitatively interpreted in terms of magnetic induced interactions between the particles in the \emph{solid} phase. The reduction of the terminal \emph{liquefaction} rate with magnetic field suggests a stabilizing effect of the magnetic field on the \emph{solid} pack. A possible interpretation of this could be the formation of chain-forces of particles sustained by attractive magnetic interactions reinforcing the global cohesion of the pack of particles. The origin of the initial acceleration of the decompaction is less clear. It may result from the vertical repulsion between parallel induced dipoles from successive layers in the initially compact \emph{solid} phase which could promote the initial destabilization of the \emph{solid}-water interface. Thus loosen, particles in the interface could then self-organize to form the stabilizing chain-forces aforementioned. Finally, the independence of particles concentration in the \emph{liquid} phase very likely reveals that when the particles are sufficiently dilute the typical inter-particle distance is so that the magnetic interaction between induced dipoles becomes negligible compared to gravity and hydrodynamic interaction. In such situation, the magnetic field then is not expected to have any significant impact on the inner structure of the dilute \emph{liquid} phase.

This final observation suggests that paramagnetic particles as the one investigated here are very likely not good candidates regarding the original motivation of the present work, aiming at using such magnetic particles in laden turbulent flows to model interacting particles with tunable interactions, as the interactions will be too weak (unless going to extremely dense seedings) compared to hydrodynamic forces. {{Regarding this aspect related to turbulent particle-laden flows, using ferromagnetic particles (rather than paramagnetic) shall allow to reach magnetic Bond numbers of the order of unity at reasonable seeding densities.}}

The study of paramagnetic particles still deserves however further developments. The proposed qualitative scenario for the decompaction process investigated in the present work suggests indeed that most of the action of the magnetic field affects primarily the \emph{solid} phase and the \emph{solid-liquid} interface, while the \emph{liquid} phase itself remains mostly insensitive to the applied field. One could then expect a clear signature of the magnetic field on the fingering (Rayleigh-Taylor like) instability occurring in the early stage of the decompaction at the interface between the \emph{solid} phase and the water. The scenario where the magnetic field would tend to increase the cohesion of the pack by magnetically induced chain-forces can be expected to act as an enhanced effective surface tension of the interface, with a possible stabilizing effect regarding the fingering instability. In the continuity of the present study, we would like to investigate the fingering instability, by a systematic investigation of the structure and the dynamics of the \emph{solid-liquid} interface (in order to extract information on the wavelength and growth rate of the instability) as a function of the amplitude of the applied magnetic field. 

\section{Acknowledgments}
This work benefited from the financial support of the Project IDEXLYON of the University of Lyon in the framework of the French program  "Programme Investissements d'Avenir" (ANR-16-IDEX-0005).
We would also like to thank Damien Leroy for helping us with the SQUID measurements to determine the magnetic susceptibility of our particles.

\medskip


\bibliographystyle{plain}
\bibliography{CFM2017}



%

\end{document}